\documentclass[aps,prb,reprint,groupedaddress]{revtex4-1}

\usepackage{graphicx,amssymb}
\usepackage{color,ulem}

\newcommand{\eqref}[1]{(\ref{#1})}
\renewcommand{\Re}{\mathop \mathrm{Re}}

\begin{document}
\title{Otto refrigerator based on a superconducting qubit: classical and quantum performance}

\author{B. Karimi}
\affiliation{Low Temperature Laboratory, Department of Applied Physics, Aalto University School of Science, P.O. Box 13500, 00076 Aalto, Finland}
\author{J. P. Pekola}
\affiliation{Low Temperature Laboratory, Department of Applied Physics, Aalto University School of Science, P.O. Box 13500, 00076 Aalto, Finland}

\date{\today}

\begin{abstract}
We analyse a quantum Otto refrigerator based on a superconducting qubit coupled to two LC-resonators each including a resistor acting as a reservoir. We find various operation regimes: nearly adiabatic (low driving frequency), ideal Otto cycle (intermediate frequency), and non-adiabatic coherent regime (high frequency). In the nearly adiabatic regime, the cooling power is quadratic in frequency, and we find substantially enhanced coefficient of performance $\epsilon$, as compared to that of an ideal Otto cycle. Quantum coherent effects lead invariably to decrease in both cooling power and $\epsilon$ as compared to purely classical dynamics. In the non-adiabatic regime we observe strong coherent oscillations of the cooling power as a function of frequency. We investigate various driving waveforms: compared to the standard sinusoidal drive, truncated trapezoidal drive with optimized rise and dwell times yields higher cooling power and efficiency.
\end{abstract}

% insert suggested PACS numbers in braces on next line
%\pacs{}

\maketitle

\section{Introduction}
Dynamical control of open systems within the framework of quantum thermodynamics is gaining increased attention. Several theoretical proposals and a few experimental ones have recently been put forward for quantum heat engines \cite{alicki1979,campisi2016,hofer2016,kosloff2014,scully2003,
quan2007,marchegiani2016,rossnagel2016,Uzdin/Kosloff} and refrigerators \cite {abah2016,brandner2016,niskanen2007,hofer2016b}. Most of the proposed engines are candidates to work in both classical and quantum regimes, but understanding the influence of quantum dynamics on their performance calls for more research \cite{brandner2016,Uzdin/Kosloff}. Different quantum systems, such as single atoms %\cite{rossnagel2016}
and superconducting circuits, are to be employed as a working substance in quantum engines, often in form of two-level systems or harmonic oscillators.
\begin{figure}[b]
\centering
\includegraphics [width=\columnwidth] {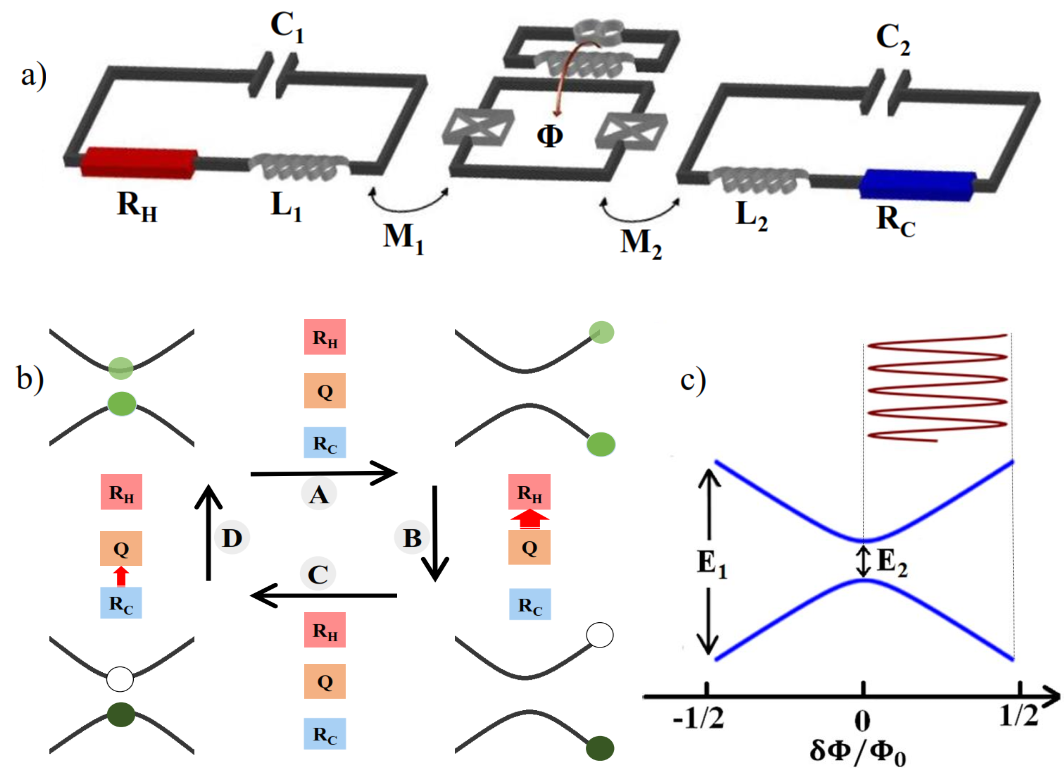}
\caption{a) Scheme of the quantum refrigerator presented. b) Thermodynamic Otto cycle of the refrigerator. c) Configuration of the two level energies of the qubit under sinusoidal driving depicted on top of the diagram.
\label{fig1.a}}
\end{figure} 

The basic Otto cycle consists of adiabatic expansion, rejection of heat at constant volume, adiabatic compression, and heat extraction at constant volume. This paper, discussing quantitatively the performance of a quantum Otto refrigerator based on a superconducting qubit is organised as follows. In Section \ref{sec2} we present the design of the refrigerator coupled to two reservoirs \cite{niskanen2007}. Using a standard quantum master equation, we analyse in Section \ref{sec3} its power in various driving frequency regimes. We present an expansion of the density matrix at low frequencies and find expressions for heat flux between the reservoirs with explicit classical and quantum contributions. Section \ref{sec4} is devoted to the discussion of different driving waveforms that yield improved performance beyond that based on the obvious sinusoidal protocol. In Section \ref{sec5}, we study the coefficient of performance of the Otto refrigerator and the effect of quantum dynamics on it. Owing to the rapid progress in superconducting qubit technology, this set-up is fully feasible for experimental implementation which will be briefly discussed in Section \ref{sec6}. 

\section{Description of the system and thermodynamic cycle}\label{sec2}

The studied quantum Otto refrigerator is schematically illustrated in Fig.~\ref{fig1.a}a. The superconducting qubit in the middle consists of a loop interrupted by Josephson junctions. It is coupled to two resonators via mutual inductances $M_1$ and $M_2$ on the left and right, and a bias circuit on top  controls the flux $\Phi$ through the loop with $q\equiv \delta\Phi/\Phi_0$. Here $\delta \Phi\equiv \Phi - \Phi_0/2$ and $\Phi_0 =h/2e$ is the superconducting flux quantum. Each resonator is a series $RLC$ circuit. Resistors $R_{\rm C}$ and $R_{\rm H}$, in general with different inverse temperatures, $\beta_1= (k_B T_{\rm C})^{-1}$ and $\beta_2= (k_B T_{\rm H})^{-1}$, are the cold and hot baths, respectively. Strictly speaking, "hot" and "cold" refer here to the resonance frequencies of the two LC-circuits, "cold" ("hot") being that with lower (higher) frequency $\omega_2$ ($\omega_1$). In general the two temperatures can take arbitrary values. In this paper we present inductive coupling of the qubit to the resonators, but this can be replaced by capacitive coupling when more appropriate.

The thermodynamic cycle of this refrigerator is sketched in Fig.~\ref{fig1.a}b and it consists of four legs labeled A - D with the following ideal properties. (A) Isentropic expansion ($q=0\rightarrow q=1/2$): the qubit is isolated from the two baths as it is not in resonance with either of the two LC-circuits, and its population is determined by the temperature of the cold resistor $R_{\rm C}$. (B) Thermalization with the hot bath: the qubit is coupled to the hot resistor $R_{\rm H}$ at $q=1/2$ and the energy flows from the qubit to the resistor. (C) Isentropic compression ($q=1/2\rightarrow q=0$): the qubit is in thermal equilibrium with the hot bath but decoupled from both the baths during the ramp. (D) Thermalization with the cold bath: the system is brought back to initial thermal state in equilibrium with the cold resistor at $q=0$. Energy in this process flows from the cold resistor to the qubit. The cycle as a whole can also be viewed as periodic alternating control of the Purcell effect of the qubit \cite{houck2008} with the two resonators.
%\begin{figure}[t!]
%  \centering
%\begin{tabular}{c}
%  \includegraphics*[width=\columnwidth,angle=0]{fig1a.png} \\
%    \includegraphics*[width=\columnwidth,angle=0]{fig1b.pdf}
%\end{tabular}
%
%%\caption{a) Scheme of the quantum refrigerator presented. b) Thermodynamic Otto cycle of the refrigerator.}
%\label{fig1.a}
%\end{figure}
\begin{figure}[t]
\centering
\includegraphics [width=8.5cm] {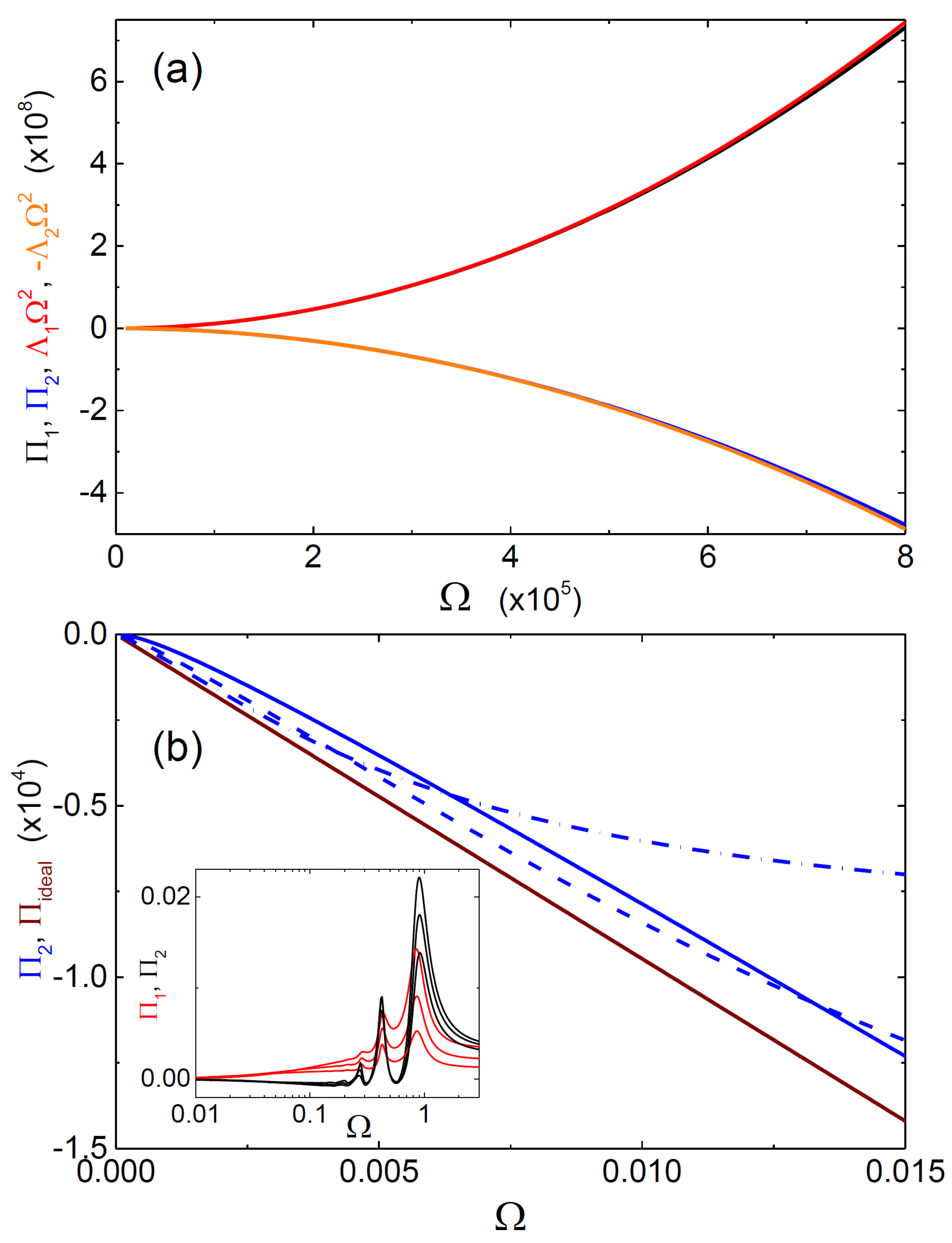}
\caption{(Color online) The powers to the hot and cold reservoirs as a function of (dimensionless) frequency $\Omega$ with chosen parameters ($k_B T_{\rm C}/E_0=k_B T_{\rm H}/E_0=0.3$, $\Delta=0.3$, $\omega_{LC,1}=2E_0 \sqrt{1/4+\Delta^2}/\hbar$, and $\omega_{LC,2}=2E_0 \Delta /\hbar$). Different operation regimes are shown separately in the plots. a) Quadratic dependence of the two powers on $\Omega$ at low frequencies with two methods (analytical and fully numerical methods). The rising parabolas are for $\Pi_1$ and the descending ones for $\Pi_2$. b) Nearly ideal Otto cycle at an intermediate frequency. The solid brown line illustrates the cooling power of an ideal Otto cycle while the other three lines are numeric cooling power when $g=g_1=g_2=1$ (solid blue line), $0.3$ (dashed line), and $0.1$ (dot dashed line). Inset of b): the non-adiabatic regime at high frequencies associated with coherent oscillations for $\Pi_1$ (red lines) and  $\Pi_2$ (black lines) with different values of $Q\equiv Q_1=Q_2$. From top to bottom $Q$=10, 30, and 100.
\label{fig2}}
\end{figure}
The Hamiltonian of the whole set-up is given by
\begin{equation}
H=H_{R_{\rm H}}+H_{R_{\rm C}}+H_{\rm cH}+H_{\rm cC}+H_{\rm Q},
\end{equation}
where $H_{R_{\rm H}}$ and $H_{R_{\rm C}}$ are the Hamiltonians of the two reservoirs, $H_{\rm Q}$ that of the qubit, and  $H_{\rm cH}$ and $H_{\rm cC}$ represent the coupling between the qubit and the corresponding reservoir. Our analysis applies to a generic superconducting qubit \cite{clarke08}: for instance, in transmon \cite{koch2007} and flux qubits \cite{mooij1999}, the two level system is formed of Josephson junctions for which $E_J/E_C\gg 1$. Here $E_J$ is the Josephson coupling energy of the junctions and $E_C$ is the Cooper pair charging energy. 
The Hamiltonian of the qubit is given by
\begin{equation} \label{hamiltonian}
H_{\rm Q}= -E_0 (\Delta \sigma_x + q \sigma_z)
\end{equation}
where $\sigma_x$ and $\sigma_z$ are the Pauli matrices, and $E_0$ is the overall energy scale of the qubit, such that the level spacing between the instantaneous eigenstates (ground state $|g\rangle$, excited state $|e\rangle$) is given by $E= 2E_0 \sqrt{q^2 +\Delta^2}$. The maximum and minimum level separations at $q=1/2$ and $q=0$ are denoted by $E_1=\hbar\omega_1$ and $E_2=\hbar\omega_2$, respectively, and $\Delta=E_2/(2E_0)$. Referring to the common transmon and flux qubits, the parameters in Eq. \eqref{hamiltonian} attain values $E_0 \sim E_J$ and $\Delta \sim E_C/E_J$.

%In order to find the transition rates of the qubit, we consider voltage noise source as a perturbation of the qubit Hamiltonian due to its coupling to the $R_j, L_j, C_j$ circuit as $E_0\sigma_z \delta \Phi _j/\Phi_0$, where $\delta \Phi _j = M_j \delta I_j$ is the fluctuation of flux in the qubit due to mutual inductance $M_j$ in response to fluctuating current $\delta I_j$ of resistor $R_j$. At frequency $\omega$, the current noise is given by $I_{j}(\omega)= V_{j}(\omega)/\{R_j+i[\omega L_j-1/(\omega C_j)]\}$. This yields the unsymmetrized noise spectrum $S_{I,j} (\omega)=\int e^{i\omega t}\langle \delta I_j(t)\delta I_j(0)\rangle dt = |R_j+i[\omega L_j-1/(\omega C_j)]|^{-2} S_{V,j} (\omega) = |\Re[ Y_j(\omega)]|^2 S_{V,j} (\omega)$, where $S_{V,j} (\omega)=2R_j \hbar\omega/(1-e^{-\beta_j\hbar\omega})$ is the voltage noise of the resistor alone, and $\Re[ Y_j(\omega)] = \{R_j[1+Q_j^2(\frac{\omega}{\omega_{LC,j}}-\frac{\omega_{LC,j}}{\omega})^2]\}^{-1}$ is the real part of the admittance of circuit $j$. Here, $\omega_{LC,j}= 1/\sqrt{L_jC_j}$ is the bare resonance angular frequency and $Q_j=\sqrt{L_j/C_j}/R_j$ the quality factor of circuit $j$. The transition rates are given by
The transition rates between the two levels of the qubit due to the two baths are given by
\begin{equation} 
\Gamma_{\downarrow , \uparrow, j}=\frac{E_0^2 M_j^2}{\hbar^2 \Phi_0^2}  \frac{\Delta^2}{q^2+ \Delta^2} S_{I,j}(\pm E/\hbar)
\end{equation}
where $S_{I,j}(\omega)=\{R_j^2[1+Q_j^2(\frac{\omega}{\omega_{LC,j}}-\frac{\omega_{LC,j}}{\omega})^2]\}^{-1} S_{V,j} (\omega)$ is the unsymmetrized noise spectrum. Here,  $\omega_{LC,j}= 1/\sqrt{L_jC_j}$ and $Q_j=\sqrt{L_j/C_j}/R_j$ are the bare resonance angular frequency and the quality factor of circuit $j$, and $S_{V,j} (\omega)=2R_j \hbar\omega/(1-e^{-\beta_j\hbar\omega})$ denotes the voltage noise of the resistor. The $+$ and $-$ sign refer to the relaxation ($\downarrow$) and excitation ($\uparrow$) of the qubit, respectively. For more details see the Appendix.

For quantitative analysis, we consider the standard master equation for the time $t$ evolution of the qubit density matrix $\rho (t)$ in the instantaneous eigenbasis  \cite{breuer,jp2016}. Ignoring pure dephasing, due to the intentionally large thermalization rate, we find the components of $\rho (t)$ as 
\begin{eqnarray} \label{e1}
&&\dot \rho_{gg}=-\frac{\Delta}{q^2+\Delta^2}\dot q \Re[\rho_{ge}e^{i\phi(t)}]-\Gamma_\Sigma \rho_{gg} +\Gamma_\downarrow \nonumber \\
&& \dot \rho_{ge} = \frac{\Delta}{q^2+\Delta^2}\dot q(\rho_{gg}-1/2)e^{-i\phi(t)} -\frac{1}{2}\Gamma_\Sigma \rho_{ge} ,
\end{eqnarray} 
where $\dot q$ is the ramp rate, $\phi(t)=\int_0^t E(t')dt'/\hbar $, $\Gamma_{\Sigma} = \Gamma_{\Sigma,1} + \Gamma_{\Sigma,2}$, $\Gamma_{\downarrow} = \Gamma_{\downarrow,1} + \Gamma_{\downarrow,2}$, and $\Gamma_{\Sigma,j} = \Gamma_{\uparrow,j}+\Gamma_{\downarrow,j}$, for $j=1,2$.

The expression of power to the resistor $j$ from the qubit is given by
\begin{eqnarray}\label{b10}
P_j && =  E(t) (\rho_{ee} \Gamma_{\downarrow,j}-\rho_{gg} \Gamma_{\uparrow,j})
\nonumber \\ &&=E(t) (\Gamma_{\downarrow,j}-\rho_{gg} \Gamma_{\Sigma,j}),
\end{eqnarray}
where $\rho_{ee}(t)=1-\rho_{gg}(t)$. The details of deriving Eq. (\ref{b10}) are presented in the Appendix. The difference between the heating power $P_1$ to reservoir $R_H$, and the cooling power $-P_2$ of reservoir $R_C$, i.e., $P_1+P_2$ equals ideally (that is with no other losses) the power that is taken from the source of the magnetic flux acting on the qubit.

\section{Different operation regimes}\label{sec3}
We identify the main operation regimes of the Otto refrigerator in three different frequency $f$ ranges: nearly adiabatic regime at low frequencies, ideal Otto cycle in the intermediate frequency regime, and non-adiabatic coherent regime at high frequencies. In Fig. \ref{fig2} we illustrate these regimes by presenting the powers to the two reservoirs in dimensionless form, $\Pi_j\equiv P_j/(E_0^2/\hbar)$, $j=1,2$, as a function of $\Omega=2\pi\hbar f/E_0$, the dimensionless frequency of the drive, for chosen parameters. We assume periodic driving $q(u)$ in (dimensionless) time $u=2\pi ft$. The powers are averaged over a cycle in steady-state under periodic driving. Below we detail the properties of the refrigerator in these three regimes.
\begin{figure}[t]
\centering
\includegraphics [width=\columnwidth] {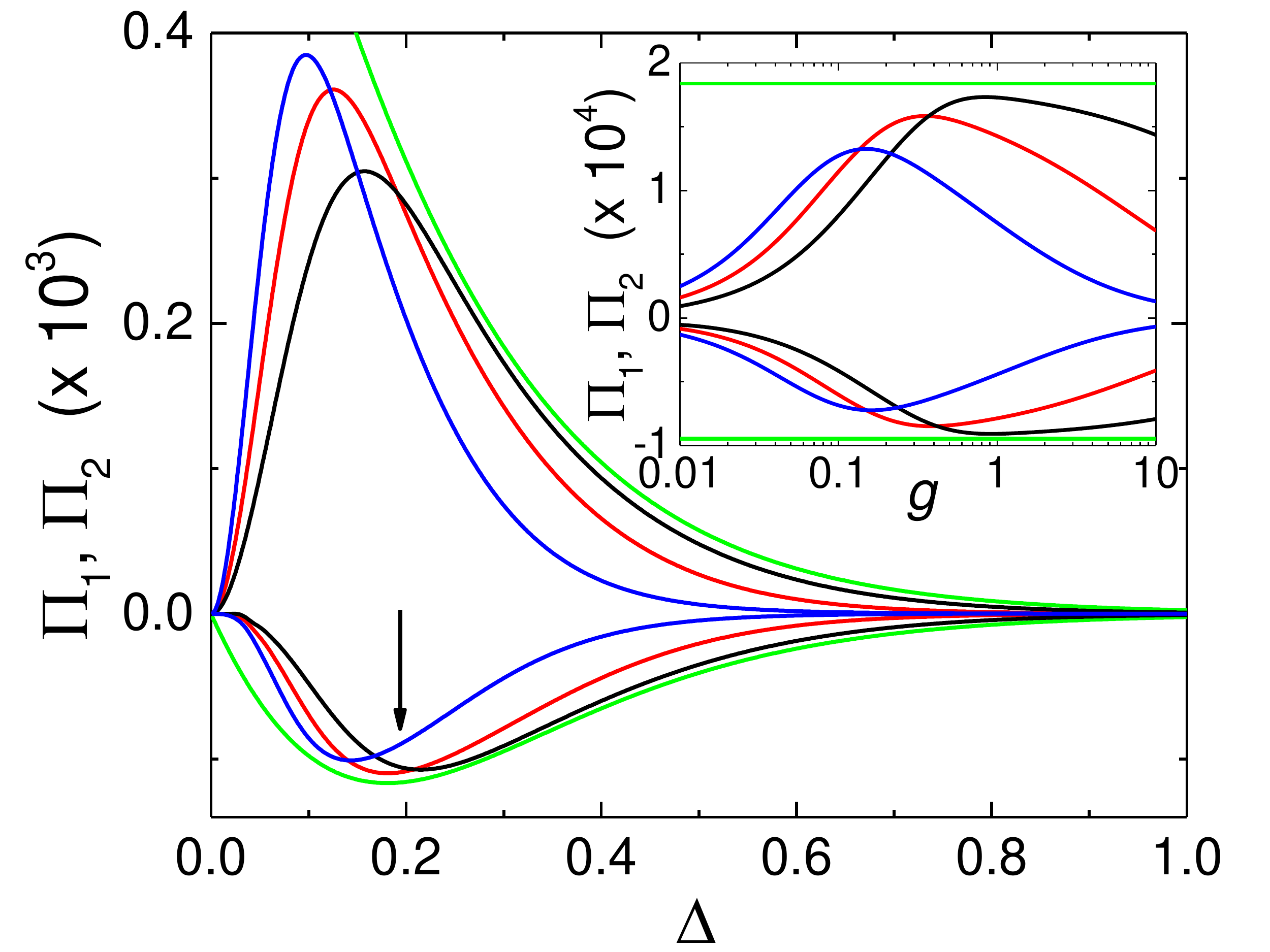}
\caption{(Color online) Dimensionless powers $\Pi_i$ as a function of $\Delta$ and $g$ (inset). The blue lines show the powers for $Q=Q_1=Q_2=10$, red for $30$, and black for $100$. The green lines display powers for the ideal Otto cycle and the arrow points to the optimal value of $\Delta$ in an ideal Otto cycle. The parameters are $k_B T_{\rm C}/E_0=k_B T_{\rm H}/E_0=0.3$, $\Omega=0.01$, and $g=g_1=g_2=1$, and for the inset $k_B T_{\rm C}/E_0=k_B T_{\rm H}/E_0=0.3$, $\Omega=0.01$, and $\Delta=0.3$.   
\label{fig3}}
\end{figure}
\\

\subsection{Nearly adiabatic regime} 
Figure \ref{fig2}a shows the cooling and heating powers of the refrigerator at low frequencies $\Omega$. We present below results for both cooling power and efficiency in the nearly adiabatic frequency range: to the best of our knowledge this regime has not been discussed quantitatively in literature in connection with a quantum four-stroke refrigerator. In order to obtain $\rho(t)$ we can here write it as an expansion in $\Omega$ as
% first we rewrite the standard master equation for density matrix (\ref{e1}) in dimensionless form as
%\begin{eqnarray} \label{e2}
%\frac{ d\rho_{gg}}{du}=-\frac{\epsilon}{q^2+\epsilon^2} \frac{dq}{du}\Re[\rho_{ge}e^{i2\kappa \int_0^u \sqrt{q^2+\epsilon^2}du'}]-\gamma_\Sigma \rho_{gg} +\gamma_\downarrow \nonumber \\
%\frac{ d\rho_{ge}}{du} = \frac{\epsilon}{q^2+\epsilon^2} \frac{dq}{du} (\rho_{gg}-1/2)e^{-i2\kappa \int_0^u \sqrt{q^2+\epsilon^2}du'} -\frac{1}{2}\gamma_\Sigma \rho_{ge}\, ,
%\end{eqnarray}
%where $u=\Omega t$, $\kappa= E_0 /(\hbar \Omega)$ and $\gamma= \frac{\hbar \kappa}{E_0} \Gamma$. In order to obtain $\rho(t)$ we can write it in an expansion form
\begin{equation} \label{b3}
\rho=\rho^{(0)}+\sum\limits_{k=1}^{\infty} \delta\rho^{(k)},
\end{equation} 
where $\rho^{(0)}$ is the density matrix at a given constant $q$, and $\delta\rho^{(k)}$ is the $k$:th order correction to it. The expression for power averaged over a cycle is given by
\begin{equation}\label{b4}
P_j= f \int_0^{1/f} dt E(t) (\Gamma_{\downarrow,j} - \Gamma_{\Sigma,j} \rho _{gg}),
\end{equation}
and for $k \ge 1$, the correction to powers can be written as
\begin{equation} \label{b5}
P_j ^{(k)}= - f  \int_0^{1/f} dt E(t)\delta \rho_{gg}^{(k)} \Gamma_{\Sigma,j}.
\end{equation}
To find $\rho^{(0)}$ in Eq. (\ref{b3}), we set $\dot{\rho}_{gg}$, $\dot{\rho}_{ge}$, and $\dot{q} $ in Eq. (\ref{e1}) equal to zero and obtain
\begin{equation}
\rho_{gg}^{(0)}=\Gamma_\downarrow /\Gamma_\Sigma\,\,\, {\rm and}\,\,\,\rho_{ge}^{(0)}=0.
\end{equation}
For equal temperature $\beta_j = \beta$ of the two reservoirs $j=1,2$, $\Gamma_\downarrow/\Gamma_\Sigma=\Gamma_{\downarrow,j}/\Gamma_{\Sigma,j}=(1+e^{-\beta  E})^{-1}$, and the power vanishes in the $0^{\rm th}$ order, $P_j^{(0)}=f \int_0^{1/f} dt E(t) (\Gamma_{\downarrow,j} - \Gamma_{\Sigma,j} \rho _{gg}^{(0)})=0$, as one would expect for fully adiabatic driving. In general for arbitrary temperatures, we find the $0^{\rm th}$ order heat flux between the two resistors, $P^{(0)} \equiv P_2^{(0)}=- P_1^{(0)}$ as an average over a "static" cycle as
\begin{widetext}
\begin{equation}
P^{(0)}=(\frac{\Delta^2 g_1 g_2}{\pi})(\frac{E_0 ^2}{\hbar})\int_0^{2\pi} du \frac{(1- e^{-\beta_2 \hbar \omega})^{-1}(e^{\beta_1 \hbar \omega} -1)^{-1}-(1-e^{-\beta_1 \hbar \omega})^{-1}(e^{\beta_2 \hbar \omega }-1)^{-1}}{g_1 [1+Q_2^2 (\frac{\omega}{\omega_{LC,2}}-\frac{\omega_{LC,2}}{\omega})^2] \coth (\frac{\beta_1 \hbar \omega}{2})+g_2 [1+Q_1^2 (\frac{\omega}{\omega_{LC,1}}-\frac{\omega_{LC,1}}{\omega})^2] \coth (\frac{\beta_2 \hbar \omega}{2})},
\end{equation}
\end{widetext}
where $g_j= \frac{4 E_0^2 M_j^2}{\hbar \Phi_0^2 R_j}$. $P^{(0)}$ does not depend on frequency and it indeed vanishes when $\beta_1 = \beta_2$. This is the heat flux that tends to counterbalance the dynamic pumping of heat in the Otto cycle, when the two temperatures are unequal. Yet due to large quality factor of the resonators, $Q_1,Q_2 \gg 1$, this contribution is typically small.

We iterate the solution in the $1^{\rm st}$ order, with the result 
\begin{equation} \label{1st}
\delta\rho_{gg}^{(1)}=-\dot{\rho}_{gg}^{(0)}/\Gamma_\Sigma 
\end{equation}
and
\begin{equation} \label{qc2}
\delta\rho_{ge}^{(1)} = \frac{\Delta}{q^2+\Delta^2}\frac{dq}{du}\, \frac{\xi_\downarrow-\xi_\uparrow}{\xi_\Sigma}\frac{e^{-i\phi}}{\xi_\Sigma - i4\sqrt{q^2+\Delta^2}} \Omega.
\end{equation}
We have defined the dimensionless rates as $\xi_i=\frac{\hbar}{E_0}\Gamma_i$. Equation \eqref{qc2} presents the quantum effects in the lowest order in $\Omega$. 
%where $u=\Omega t$. By inserting $\dot{\rho}_{gg}^{(0)}=\frac{d\rho_{gg}^{(0)}}{du}\frac{E_0}{\hbar}\Omega$ in equatin (\ref{b5}) we have
%\begin{equation}\label{b7}
%\Pi_j^{(1)}=\frac{P_j^{(1)}}{E_0^2/\hbar}=\frac{1}{\pi} \int_0^{2\pi} du \sqrt{q^2 + \Delta^2} \frac{d\rho_{gg}^{(0)}}{du}\frac{\Gamma_{\Sigma,j}}{\Gamma_\Sigma } 
%\end{equation}
%Making a change in variable from $u$ to $q$ and using $du=\frac{1}{dq/du}dq$, the equation \eqref{b7} becomes
%\begin{equation}
%\Pi_j^{(1)}=\frac{1}{\pi} \int_{q_i}^{q_f} dq \sqrt{q^2 + \Delta^2} \frac{d\rho_{gg}^{(0)} (q)}{du}\frac{\Gamma_{\Sigma,j}(q)}{\Gamma_\Sigma (q) }.
%\end{equation}
%In one cycle, because $q_i=q_f$, so 
Irrespective of the waveform we have $P_j^{(1)}=0$ (see Appendix for details).
%\begin{eqnarray}
%P_2 ^{(1)}= \int_0^{\pi} du \Delta E(u)\frac{\gamma_{\Sigma,2} (u)}{\gamma_\Sigma (u)} \frac{d\rho_{gg}^{(0)}(u)}{du}\nonumber \\+ \int_{\pi}^{0} (-d\nu) \Delta E(2 \pi - \nu)\frac{\gamma_{\Sigma,2} (2 \pi - \nu)}{\gamma_\Sigma (2 \pi - \nu)}(- \frac{d\rho_{gg}^{(0)}(\nu)}{d\nu} )
%\end{eqnarray}
%because of the symmetric drive $q(2 \pi - \nu)= q(\nu)$ then we have $ \Delta E(2 \pi - \nu)= \Delta E(\nu) $,$\gamma_{\Sigma,2} (2 \pi - \nu)= \gamma_{\Sigma,2} (\nu)$ and $\gamma_\Sigma (2 \pi - \nu)= \gamma_\Sigma (\nu)$,
%\begin{eqnarray}
%P_2 ^{(1)}= \int_0^{\pi} du \Delta E(u)\frac{\gamma_{\Sigma,2} (u)}{\gamma_\Sigma (u)} \frac{d\rho_{gg}^{(0)}(u)}{du}\nonumber\\ - \int_{0}^{\pi} (d\nu) \Delta E( \nu)\frac{\gamma_{\Sigma,2} ( \nu)}{\gamma_\Sigma (\nu)}(\frac{d\rho_{gg}^{(0)}(\nu)}{d\nu} )=0
%\end{eqnarray}
%so for symmetric drive only we have $P_2 ^{(1)}=0$ , but temperatures of the baths do not need to be equal in this case.\\
The first non-vanishing contribution to the powers comes from the second order diagonal element
\begin{equation}\label{b6}
\delta\rho_{gg}^{(2)}=\frac{\frac{d^2 \rho_{gg}^{(0)}}{du^2}}{\xi_\Sigma^2}- \frac{\frac{d\rho_{gg}^{(0)}}{du}\frac{d\xi_\Sigma}{du}}{\xi_\Sigma^3}-\frac{\Delta}{q^2+\Delta^2}\frac{dq}{du} \frac{1}{\xi_\Sigma}{\rm Re}(\delta\rho_{ge}^{(1)} e^{i\phi}) \Omega.
\end{equation}
The third term of Eq. (\ref{b6}) is the pure quantum correction of $\rho_{gg}$. In dimensionless form, we have then
\begin{equation} \label{pilambda}
\Pi_j^{(2)}=\Lambda_j\Omega^2.
\end{equation}
We can separate the classical contribution $\Lambda_{j,{\rm CL}}$ and the quantum correction $\delta\Lambda_{j,{\rm Q}}$ of $\Lambda_j$, such that $\Lambda_j=\Lambda_{j,{\rm CL}}+\delta\Lambda_{j,{\rm Q}}$, where
\begin{widetext}
\begin{equation} 
\Lambda_{j,{\rm CL}}=-\frac{1}{\pi}  \int_0^{2\pi} du \sqrt{q^2 + \Delta ^2}( \frac{\frac {d^2\rho_{\rm eq, gg}}{du^2}}{\xi_\Sigma ^2} - \frac{(\frac{d\rho_{\rm eq, gg}}{du})(\frac{d\xi_\Sigma}{du})}{\xi_\Sigma ^3})\xi_{\Sigma,j}
\end{equation}
and 
\begin{equation} \label{Lambdas}
\delta \Lambda_{j,{\rm Q}} = \frac{1}{\pi}\int_0^{2\pi} du \frac{\Delta^2}{(q^2+\Delta^2)^{3/2}}(\frac{dq}{du})^2\, \frac{(\xi_\downarrow-\xi_\uparrow)\xi_{\Sigma,j}}{\xi_\Sigma [\xi_\Sigma^2 + 16(q^2+\Delta^2)]}.
\end{equation}
\end{widetext}
We observe that based on Eq. \eqref{pilambda}, the energy transferred in a cycle reads $P_j/f = 2\pi \Lambda_j hf$. Yet the dimensionless prefactor  $2\pi \Lambda_j$ is system-dependent, in particular it depends inversely on the coupling $g$. This dependence is vivid in Fig. \ref{fig2}b if one zooms the very low $\Omega$ regime for different values of $g$.

In the quadratic regime, the total powers on the two resistors $j$ can be written for arbitrary temperatures as 
\begin{equation}
\Pi_j= (-1)^j \Pi^{(0)} +{\Lambda}_j \Omega^2 .
\end{equation}
The results of the fully numerical calculation are shown together with the semi-analytic quadratic result $\Lambda_j \Omega^2$ in Fig. \ref{fig2}a for the equal temperature case. The two results are nearly indistinguishable.

It is interesting to note that the coherent effects via $\delta \Lambda_{j,{\rm Q}}$ increase the dissipation unconditionally. This is because the integrand of the quantum correction in Eq. \eqref{Lambdas} is strictly non-negative; in particular all the rates $\xi_i$ are positive and moreover, $\xi_\downarrow > \xi_\uparrow$. 

\subsection{Intermediate frequencies (Otto cycle)} 
In the intermediate regime, as shown in Fig. \ref{fig2}b, the cooling power $-P_2$ is approximately linear in frequency with a slope given below in Eq. \eqref{b8}. This behaviour corresponds to the ideal Otto cycle. To find the powers $P_1,P_2$, we assume that the qubit thermalizes at both $q=0$ and $q=1/2$, and that the population of the qubit does not change between the two extremes of the cycle. At  $q=1/2$, the qubit population is $\rho_{gg}=1/(1+e^{-\beta_2\hbar \omega_2})$. When brought to $q = 0$, $\rho_{gg}$ ideally attains the value $\rho_{gg}=1/(1+e^{-\beta_1\hbar\omega_1})$ when interacting with $R_{\rm C}$. In this process energy is transferred from resistor $R_{\rm C}$ to the qubit, ideally with power $-P_2$ and from the qubit to resistor $R_{\rm H}$ with power $P_1$ given by \cite{niskanen2007}
\begin{eqnarray} \label{b8}
&&P_1 = +\frac{\hbar\omega_1}{2}[\tanh(\frac{\beta_1\hbar\omega_1}{2})-\tanh(\frac{\beta_2\hbar\omega_2}{2})]f,\nonumber\\
&&P_2 = -\frac{\hbar\omega_2}{2}[\tanh(\frac{\beta_1\hbar\omega_1}{2})-\tanh(\frac{\beta_2\hbar\omega_2}{2})]f.
\end{eqnarray}
These powers depend critically on the energy separation at $q=0$. We maximize the cooling power $-P_2$ of Eq. (\ref{b8}) with respect to $\omega_2$ keeping other parameters constant, obtaining
\begin{equation}
\tanh(\beta_2 E_0 \Delta) + \frac{\beta_2 E_0 \Delta}{\cosh^2(\beta_2 E_0 \Delta)} - \tanh (\beta_1 E_0 \sqrt{\frac{1}{4}+\Delta^2})=0.
\end{equation}
We assume that the gap at $q=1/2$ is large enough such that we can set $\tanh (\beta_1 E_0 \sqrt{\frac{1}{4}+\Delta^2})\simeq 1$. This yields the equation $2x-e^{-2x} -1=0$ for $x=\beta_2 E_0 \Delta$, with $x=0.6392...$ as the solution. 
%By selecting the parameter $\theta_2$ to be $\frac{10}{3}$, the optimum value for $\Delta$ will be
%\begin{equation}
%\Delta_{\rm optimum}=0.192
%\end{equation}
Numerically obtained powers to the two resistors as a function of  $\Delta$ and $g\equiv g_1=g_2$ (inset) are shown in Fig. \ref{fig3} for typical parameters. These figures are plotted for different quality factors of the $RLC$ circuits. The vertical arrow indicates the optimal point $x = 0.6392...$ obtained above. It is vivid that the maximum value of cooling power shifts towards higher values of $\Delta$ and $g$ when increasing $Q$, and for $Q=100$, the powers are very close to those of the ideal Otto cycle [Eq. \eqref{b8}] at this value of frequency ($\Omega =0.01$).
\subsection{Non-adiabatic coherent regime} 
At high frequencies coherent oscillations of the qubit are reflected in the powers as seen in the inset of Fig. \ref{fig2}b. The oscillatory regime essentially spans frequencies from $E_2/(2\pi\hbar)$ to $E_1/(2\pi\hbar)$. In this frequency range, the population of the qubit in the adiabatic legs of the cycle does not remain constant due to driving-induced coherent oscillations. At still higher frequencies, both powers are positive (dissipative) and almost constant. Lower $Q$ means more dissipation in general, explaining the relative results of $\Pi_j$ in the figure for different quality factors. One needs to bear in mind, however, that our analysis based on instantaneous eigenstates is not rigorous at these high frequencies \cite{jp2010,salmilehto2011} that may also exceed the bath correlation time in practise.
\begin{figure}[t]
\centering
\includegraphics [width=\columnwidth] {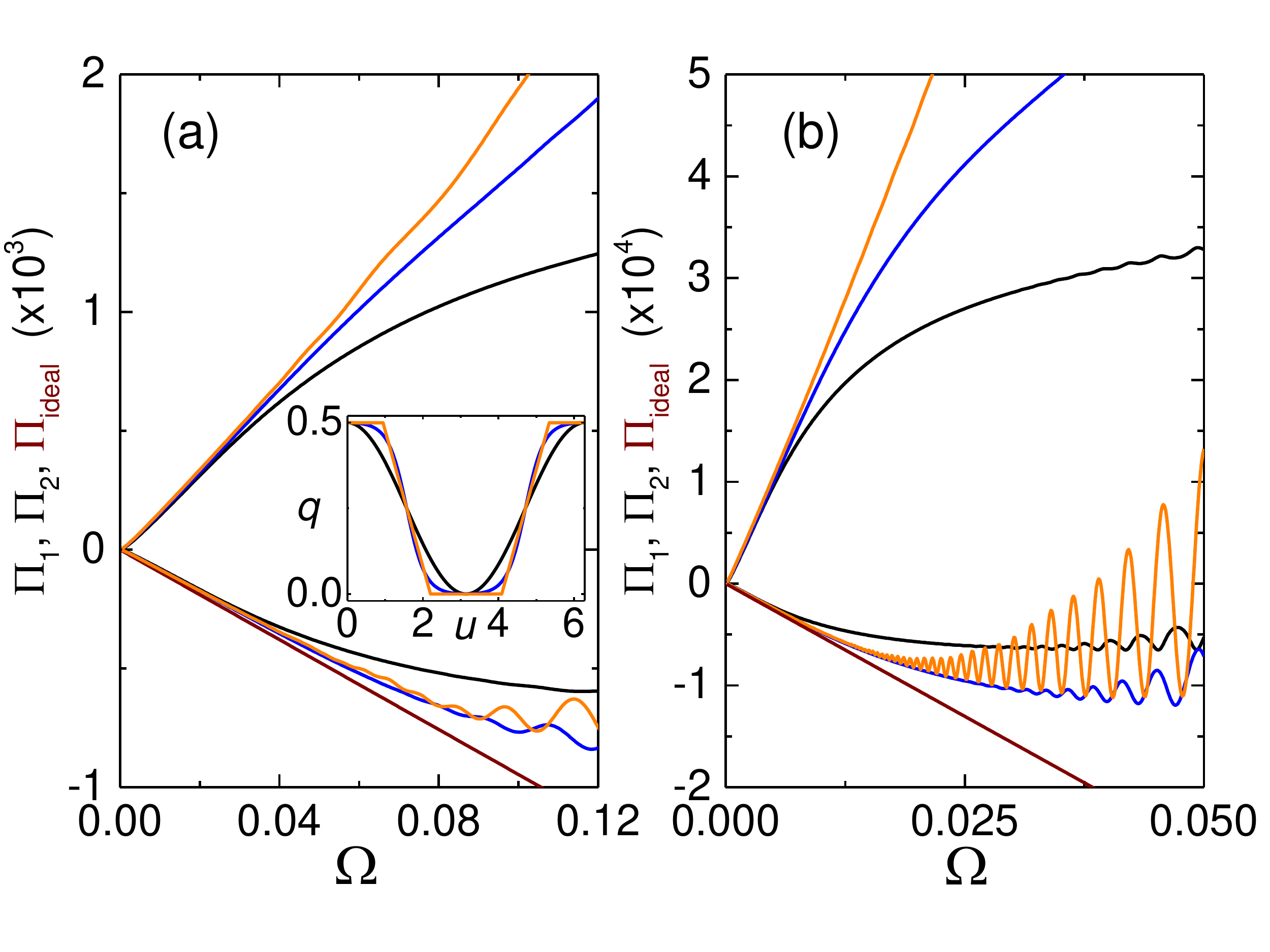}
\caption{(Color online) The variation of powers $\Pi_1$ (ascending curves) and $\Pi_2$ (descending curves) as a function of frequency in the intermediate regime. Black lines display powers for sinusoidal drive, light brown for trapezoidal drive, blue for truncated trapezoidal drive, and dark brown for ideal Otto cycle (for $\Pi_2$), with $g=g_1=g_2=1$ and $Q=Q_1=Q_2= 30$. a) Equal temperature of the two reservoirs $(k_B T_{\rm C} /E_0=k_B T_{\rm H} /E_0=0.3)$ and $\Delta=0.3$. b) Different bath temperatures $(k_B T_{\rm H} /E_0=2k_B T_{\rm C} /E_0=0.3)$ and $\Delta=0.12$. Inset in a: The considered driving waveforms; sinusoidal, trapezoidal, and truncated trapezoidal.
\label{fig4}}
\end{figure}

\section{Different driving waveforms}\label{sec4}
In assessing the influence of the driving waveform on the cooling power and efficiency of the refrigerator, we apply sinusoidal $q(u)=  \frac{1}{4}(1+ \cos u)$, trapezoidal (specifically with symmetric form consisting of rising sections of 20\% of the cycle time each, and plateaus of 30\% duration each), and truncated trapezoidal $q(u)= \frac{1}{4}[1+ \tanh(a \cos u)/\tanh a]$, specifically with $a=2$. These rising times and the particular value of $a$ yield nearly optimal performance under the conditions of our numerical simulations for the two latter waveforms. See the inset of Fig.~\ref{fig4}a for the illustration of the three protocols.

\begin{figure}[t]
\centering
\includegraphics [width=\columnwidth] {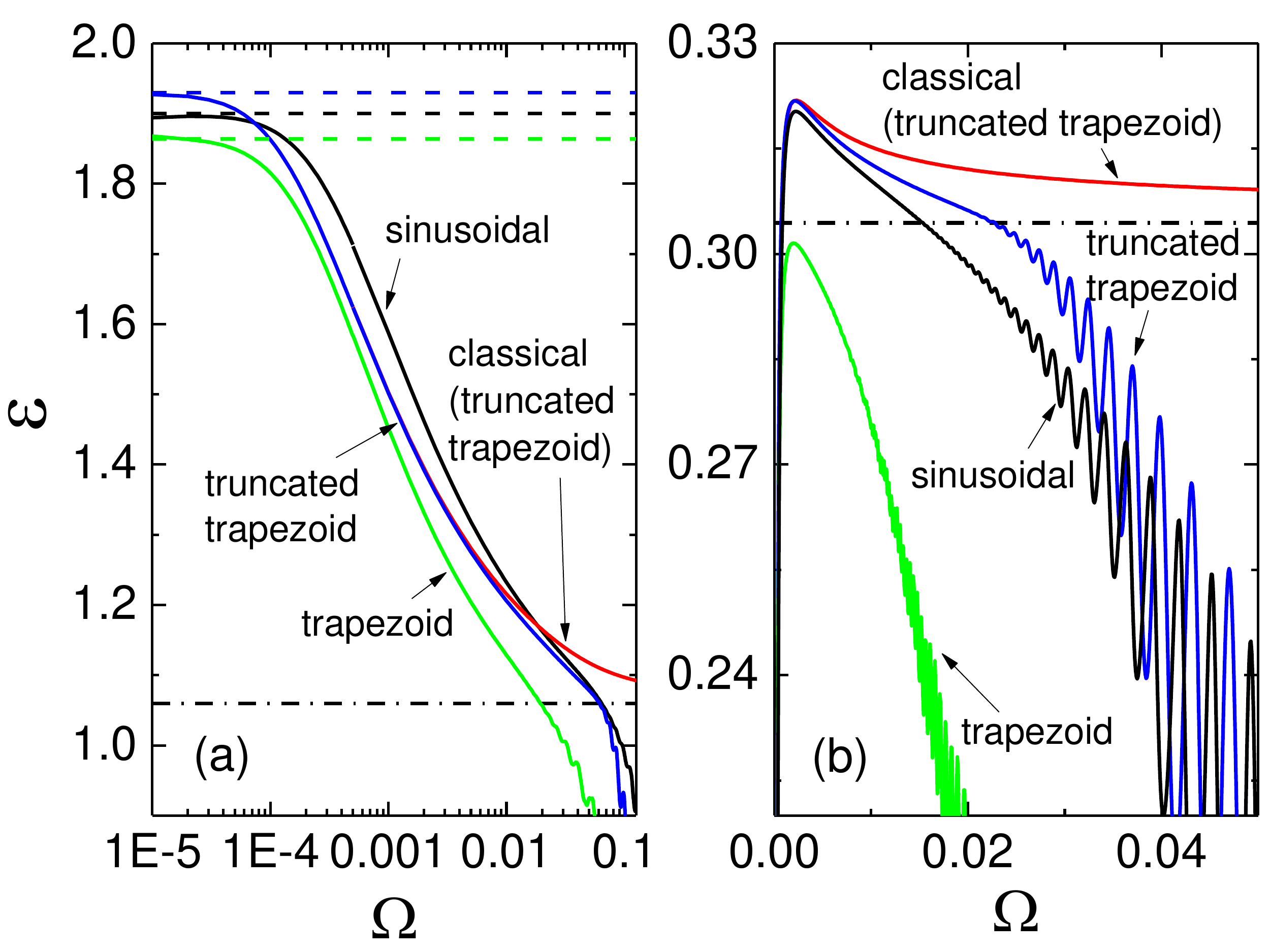}
\caption{(Color online) Dependence of the coefficient of performance $\epsilon$ on frequency for different drives as indicated by the names with arrows. Dot dashed lines correspond to ideal efficiency of the Otto refrigerator, $\epsilon_{\rm ideal}$, with $Q=Q_1=Q_2= 30$ and $g=g_1=g_2=1$ for both panels. The horizontal dashed lines represent analytical results of efficiency for different drives in the low frequency regime based on Eq. (\ref{b11}). The parameters are a) $k_B T_{\rm H}/E_0 =k_B T_{\rm C}/E_0 =0.3$ and $\Delta=0.3$, b) $k_B T_{\rm H}/E_0 =0.3, k_B T_{\rm C}/E_0 = 0.15$ and $\Delta=0.12$. 
\label{fig5}}
\end{figure}
The obtained dimensionless powers $\Pi_1$ and $\Pi_2$ as a function of frequency are displayed in Fig. \ref{fig4}. The data in Fig. \ref{fig4}a,b are for equal and unequal temperatures of the two reservoirs, $\beta_1=\beta_2$ and $2\beta_1=\beta_2$, respectively. At equal temperatures we can obtain higher cooling power with trapezoidal and truncated trapezoidal drives than with sinusoidal drive, while in the case of unequal temperatures, the highest values of cooling power are obtained with truncated trapezoidal drive. The inferior performance of the sinusoidal drive stems from the short available thermalization times at $q=0,1/2$, whereas the large dissipation $\Pi_1$ with the trapezoidal drive is likely to originate from the abrupt changes of the slope of this waveform \cite{jp2016}.

\section{Efficiency of the Otto refrigerator}\label{sec5}
The efficiency of a refrigerator is defined by the coefficient of performance $\epsilon$ as
\begin{equation}
\epsilon=-\mathcal Q_2/W,
\end{equation}
where $\mathcal Q_2=\int dt \dot{\mathcal Q}_2 (t)$ is the heat deposited to the cold bath in a steady state cycle (the integral is extended over such a cycle), and $W$ is the work done to achieve this. If we ignore the parasitic losses in producing the flux drive of the qubit (which can be made arbitrarily small in principle), we have $W=\int dt (\dot{\mathcal Q}_1 (t)+\dot{\mathcal Q}_2 (t))$. We have then
$\epsilon=-\mathcal Q_2/(\mathcal Q_1 +\mathcal Q_2)$.
There are two reference values to be considered. One is the Carnot efficiency of a refrigerator, given by $\epsilon_{\rm C}=1/(T_H/T_C -1)$, which can not be exceeded. Another one is the ideal $\epsilon$ of the Otto refrigerator, which turns out to be
\begin{equation}
\epsilon_{\rm ideal}=\frac{1}{\omega_1/\omega_2 -1}
\end{equation}
according to Eqs. (\ref{b8}). Based on our result of Eq. \eqref{pilambda}, we introduce $\epsilon_p$ for quadratic low frequency regime given by 
\begin{equation}\label{b11}
\epsilon_p=\frac{1}{\Lambda_1/|\Lambda_2| -1}
\end{equation}
for the equal temperature case. Numerical results on $\epsilon$ as a function of $\Omega$ for different waveforms are presented by solid lines in Fig. \ref{fig5}. It is evident in Fig. \ref{fig5}a that at equal temperatures $\beta \equiv \beta_1=\beta_2$, the truncated trapezoidal drive has the highest efficiency among the three driving protocols at low frequencies, but all of them are, somewhat surprisingly, higher than $\epsilon_{\rm ideal}$ shown by the dash-dotted horizontal line. Naturally the Carnot efficiency exceeds all other efficiencies in the figure: in a $\epsilon_{\rm C} =\infty$, and in b $\epsilon_{\rm C} = 1$. Thus we see that our system reaches high efficiency at low frequencies, which is consistent with general expectations of thermodynamics towards the adiabatic limit. The dashed lines illustrate the semi-analytic result of $\epsilon_p$ for different drives. These results are fully consistent with numerical ones at low frequency. For unequal bath temperatures, $\beta_2=2\beta_1$, in Fig. \ref{fig5}b, we have a similar hierarchy among the three waveforms, but with these parameters the (abrupt) trapezoidal drive does not even reach the efficiency of the ideal Otto cycle at any frequency. The rising part at low frequencies is due to the finite $P^{(0)}$ at unequal temperatures. For reference, the results ignoring quantum effects, solving the corresponding rate equation $\dot \rho_{gg}=\Gamma_\downarrow -\Gamma_\Sigma \rho_{gg}$ with truncated trapezoidal drive are shown in a and b by the red line. These results lie above any other curve, which is consistent with what we obtained for the quantum correction of $\Lambda_j$ in the quadratic low frequency regime. That is, the numerical result supports the observation that quantum corrections decrease the efficiency of the quantum Otto refrigerator, in agreement with the general linear response results in \cite{brandner2016}.

\section{Experimental feasibility}\label{sec6}

Finally we give few remarks on experimental parameters. The energy scale of a typical superconducting qubit is of order $E_0/k_B \sim 1$ K \cite{clarke08}. With realistic mutual inductances $M_i$, values for coupling up to $g_i \sim 1$ can be achieved with proper design \cite{niskanen2007}. The quality factors in the range presented in this manuscript can also be achieved, since a typical $\sqrt{L/C}$ impedance is of order $10^2$ $\Omega$, and a metallic resistor can have values in the range of $\sim 1$ $\Omega$. With these values, the presented numerical graphs are feasible, and the power $E_0^2/\hbar \sim 1$ pW and frequency $E_0/2\pi\hbar \sim$ 1 GHz scales should lead to experimentally observable heat fluxes (several fW) \cite{koski15} at feasible operation frequencies (100 MHz) \cite{clarke08}.
\\

In conclusion, we have investigated theoretically quantum Otto refrigerator using a generic superconducting qubit. Explicit expressions for quadratic dependence of power on low frequencies were obtained. We show that the quantum dynamics inevitably decreases, as compared to the corresponding fully classical case, both the cooling power and the efficiency of the refrigerator, but it leads to interesting oscillatory behaviour of power versus frequency. Different driving waveforms were studied, and we found that the coefficient of performance $\epsilon$ can exceed that of the ideal Otto refrigerator at low frequencies.

We thank Dmitry Golubev, Michele Campisi, Rosario Fazio, Kay Brandner, Alberto Ronzani and Jorden Senior for discussions. Financial support from the Academy of Finland (grants 272218 and 284594) is gratefully acknowledged.

\section*{Appendix}
\setcounter{section}{1}
\setcounter{equation}{0}
\setcounter{figure}{0}
\setcounter{table}{0}
\renewcommand{\theequation}{A\arabic{equation}}
\renewcommand{\thefigure}{A\arabic{figure}}
%\renewcommand{\bibnumfmt}[1]{[A#1]}
%\renewcommand{\citenumfont}[1]{A#1}
%\section{Section 1}
We present here the derivation of the expressions for the transition rates and power to each resistor due to its coupling to the qubit [Eq. \eqref{b10}] and calculation of the (vanishing) first order contributions $\Pi_j^{(1)}$ [Eq. \eqref{b5} for $k=1$]. 

\subsection{Transition rates and powers}
The Golden Rule transition rates between the instantaneous eigenstates due to the baths (resistors $j$ in Fig. \ref{fig1.a}a) are given by
\begin{equation} \label{A1}
\Gamma_{\uparrow,\downarrow,j}=\frac{1}{\hbar^2}|\langle g| \frac{\partial H}{\partial \Phi}|e\rangle|^2 M_j^2 S_{I,j}(\pm E/\hbar),
\end{equation}
where the $\pm$ signs correspond to relaxation and excitation, respectively, and $S_{I,j}(\pm \omega)$ is the unsymmetrized noise spectrum of the qubit which is 
\begin{eqnarray}
S_{I,j} (\omega)&&=\int e^{i\omega t}\langle \delta I_j(t)\delta I_j(0)\rangle dt \nonumber\\
&&= \{R_j^2[1+Q_j^2(\frac{\omega}{\omega_{LC,j}}-\frac{\omega_{LC,j}}{\omega})^2]\}^{-1} S_{V,j} (\omega)\nonumber\\ &&= R_j^{-1}\Re[ Y_j(\omega)] S_{V,j} (\omega).
\end{eqnarray}
Here, $S_{V,j} (\omega)=2R_j \hbar\omega/(1-e^{-\beta_j\hbar\omega})$ is the voltage noise of the resistor alone, and $\Re[ Y_j(\omega)] = \{R_j[1+Q_j^2(\frac{\omega}{\omega_{LC,j}}-\frac{\omega_{LC,j}}{\omega})^2]\}^{-1}$ is the real part of admittance of circuit $j$, $\omega_{LC,j}= 1/\sqrt{L_jC_j}$ and $Q_j=\sqrt{L_j/C_j}/R_j$. 
By using Eq. \eqref{hamiltonian} for the Hamiltonian of the qubit, we have $\frac{\partial H}{\partial \Phi}= -\frac{E_0 \sigma_z}{\Phi_0}$, and in order to calculate $\langle g| \sigma_z|e\rangle$, we consider the eigenvectors of the Hamiltonian, $|g\rangle =\begin{pmatrix} {\cos(\theta/2) \ \sin(\theta/2)} \end{pmatrix}$ and $|e\rangle =\begin{pmatrix} {-\sin(\theta/2) \ \cos(\theta/2)} \end{pmatrix}$. Here the angle $\theta$ is given by $\tan\theta=\Delta /q$. Then for Eq. \eqref{A1} we have
\begin{equation}\label{A2}
\Gamma_{\uparrow,\downarrow,j}=\frac{E_0^2 M_j^2}{\hbar^2 \Phi_0^{2}} \frac{\Delta^2}{q^2+\Delta^2} S_{I,j}(\pm E/\hbar).
\end{equation}
Equation \eqref{A2} yields the transition rates for a generic superconducting qubit with the Hamiltonian \eqref{hamiltonian}. For instance in the flux qubit, the factor $E_0/\Phi_0$ equals $I_p$, the persistent circulating current in the qubit loop \cite{niskanen2007}. In order to evaluate powers $P_j$, we first calculate the operator for the heat current from the resistors to the qubit as
\begin{equation}\label{A3}
\dot{H}_{\rm Q}= -\frac{i}{\hbar} [H_{\rm Q}, H_{\rm cC}+H_{\rm cH}].
\end{equation}
By inserting $H_{\rm cC}= \frac{E_0}{\Phi_0} M_1 \delta I_1(t)\sigma_z$ and $H_{\rm cH}=\frac{E_0}{\Phi_0} M_2 \delta I_2 (t)\sigma_z$ in \eqref{A3} and with $[\sigma_i, \sigma_j ]= 2i\varepsilon_{ijk} \sigma_k,$ we have
\begin{equation}
\dot{H}_{\rm Q}=2\frac{E_0^2 \Delta }{\hbar \Phi_0} [ M_1  \delta I_1(t) + M_2 \delta I_2(t)]\sigma_y.
\end{equation}
Now, in the interaction picture, with operators $\mathcal O_I(t) =e^{iH_{\rm Q}t/\hbar}\mathcal O e^{-iH_{\rm Q}t/\hbar}$, we have the expectation value of the operator $-\dot H_{\rm Q} $, i.e., the heat deposited to the two resistors by the qubit in linear response (Kubo formula) as
\begin{equation} \label{A4}
P=-\langle \dot H_{\rm Q} \rangle=\frac{i}{\hbar}\int_{-\infty}^t dt' \langle[\dot H_{{\rm Q},I}(t),H_{\rm c}(t')]\rangle,
\end{equation}
where $H_c = H_{{\rm cC},I}+H_{{\rm cH},I}$. Substituting the expressions $\langle g|\sigma_z|e\rangle = \Delta/\sqrt{q^2+\Delta^2}$, $\langle g|\sigma_y|e\rangle =i$, $\langle g|\sigma_y|g\rangle = \langle e|\sigma_y|e\rangle =0$, $\langle g|e^{iH_{\rm Q}t/\hbar}\sigma_y \sigma_z e^{-iH_{\rm Q}t'/\hbar}|g\rangle = \langle e|e^{iH_{\rm Q}t'/\hbar}\sigma_z \sigma_y e^{-iH_{\rm Q}t/\hbar}|e\rangle =i e^{-i\omega (t-t')}\Delta/\sqrt{q^2+\Delta^2}$, and $\langle e|e^{iH_{\rm Q}t/\hbar}\sigma_y \sigma_z e^{-iH_{\rm Q}t'/\hbar}|e\rangle = \langle g|e^{iH_{\rm Q}t'/\hbar}\sigma_z \sigma_y e^{-iH_{\rm Q}t/\hbar}|g\rangle = -i e^{i\omega (t-t')}\Delta/\sqrt{q^2+\Delta^2}$ in Eq. (\ref{A4}), we have $P=P_1+P_2$, where
\begin{equation}
P_j = E(t) \big{(}\rho_{ee} \Gamma_{\downarrow,j}-\rho_{gg} \Gamma_{\uparrow,j}\big{)}.
\end{equation}

\subsection{Vanishing first order contribution to powers}

The first order in $\Omega$ contribution to the powers can be written as
\begin{eqnarray} \label{A5}
P_j ^{(1)}= - f  \int_0^{1/f} dt  E(t)\delta \rho_{gg}^{(1)} \Gamma_{\Sigma,j}\nonumber\\= \frac{E_0}{\pi} \int_{0}^{2 \pi} du \sqrt{q^2 + \Delta^2}\dot{\rho}_{gg}^{(0)} \frac{\Gamma_{\Sigma,j}}{\Gamma_\Sigma } 
\end{eqnarray}
with the help of Eq. \eqref{1st}. Here $u=2\pi f t$. By inserting $\dot{\rho}_{gg}^{(0)}=\frac{d\rho_{gg}^{(0)}}{du}\frac{E_0}{\hbar}\Omega$ in Eq. (\ref{A5}) we have
\begin{equation}\label{A6}
\Pi_j^{(1)}=\frac{P_j^{(1)}}{E_0^2/\hbar}=\frac{1}{\pi} \int_0^{2\pi} du \sqrt{q^2 + \Delta^2} \frac{d\rho_{gg}^{(0)}}{du}\frac{\Gamma_{\Sigma,j}}{\Gamma_\Sigma }. 
\end{equation}
With a change of integration variable from $u$ to $q$ and using $du=\frac{1}{dq/du}dq$, Eq. \eqref{A6} becomes
\begin{equation}
\Pi_j^{(1)}=\frac{1}{\pi} \int_{q_i}^{q_f} dq \sqrt{q^2 + \Delta^2} \frac{d\rho_{gg}^{(0)}}{du}\frac{\Gamma_{\Sigma,j}}{\Gamma_\Sigma}.
\end{equation}
In cyclic operation, the initial and final values of $q$ are equal, $q_i=q_f$, and irrespective of the waveform we have $\Pi_j^{(1)}=0$.

\end{document}